\pdfoutput=1
\documentclass{article}
\usepackage{spconf}

\usepackage{amsfonts,amsmath}
\usepackage{xcolor}
\usepackage{graphicx}
\usepackage{cite}
\usepackage[title]{appendix}
\usepackage[utf8]{inputenc}
\usepackage{enumitem}
\usepackage[linesnumbered,algoruled,boxed,lined]{algorithm2e}
\usepackage{subfigure}

\graphicspath{{figures/}}

\usepackage{tikz}
\usetikzlibrary{positioning}
\usetikzlibrary{matrix,calc}

\let\oldbibliography\thebibliography
\renewcommand{\thebibliography}[1]{%
  \oldbibliography{#1}%
  \setlength{\itemsep}{-2pt}%
}

\title{Handling Background Noise in Neural Speech Generation}

\name{\begin{tabular}{c}
    Tom Denton,$^{1}$
    Alejandro Luebs,$^{1}$
    Michael Chinen,$^{1}$
    Felicia S. C. Lim,$^{1}$
    Andrew Storus,$^{1}$ \\
    Hengchin Yeh,$^{1}$
    W. Bastiaan Kleijn,$^{1,2}$
    Jan Skoglund$^{1}$
    \vspace{-1mm}
\end{tabular}}

\address{
    $^{1}$Google LLC, San Francisco, CA \hspace{0.5em}
    $^{2}$Victoria University of Wellington, NZ\\
}

\begin{document}
\ninept{}
\maketitle{}
\begin{abstract}
Recent advances in neural-network based generative modeling of speech has shown  
great potential for speech coding. However, the performance of such models
drops when the input is not clean speech, e.g., in the presence of background noise, preventing its use in
practical applications. In this paper we examine the reason and discuss methods to overcome this issue.
Placing a denoising preprocessing stage when extracting features and target clean speech during training
is shown to be the best performing strategy.
\end{abstract}

\section{Introduction}
\label{s:introduction}
Autoregressive neural synthesis systems are based on the idea that the speech signal's
probability distribution can be formulated as a scalar autoregressive structure, where the 
probability of each speech sample $s_t$ is conditioned on previous samples and 
a set of conditioning features (spectral information, pitch, etc.), $\theta_t$,
\begin{equation}
p(s_t |s_{t-1}, s_{t-2}, \ldots, \theta_t).
\end{equation}

The paradigm was first introduced for text-to-speech using the WaveNet \cite{oord2016wavenet} architecture. 
Soon thereafter WaveNet was shown beneficial also for low bit rate speech coding   
in \cite{kleijn2018wavenet}, which was the first coder using neural generative synthesis.
Another example of a codec using WaveNet as synthesis generation is \cite{garbacea2019low}. 
Since then other autoregressive generators with lower complexity have been introduced, and codecs based on such
are for example \cite{klejsa2019high}, based on SampleRNN \cite{mehri2016samplernn}, and
\cite{valin2019lpcnetcoded}, based on WaveRNN 
\cite{kalchbrenner2018efficient}. 

However, the reproduction of real-world speech signals by generative models is still a challenge. 
Coding real-world speech signals with a neural vocoder requires solving a number of problems simultaneously. 
Foremost is handling of 
background noise, but a successful system must also be able to reliably reproduce speech from arbitrary speakers 
using a low bit rate input stream, ideally with a model small and fast enough to run on a standard smartphone.
High quality has been achieved only for clean input signals and, to date, no coding performance has been 
reported for noisy speech.

The difficulty of coding noisy signals can perhaps be explained by the signal structure, where the signal to be coded 
is the sum of a clean speech signal and an interfering signal.
The autoregressive architecture is a good match for the structure
of speech, but the addition of a second signal removes this match. We
note that this phenomenon is well-known in linear modeling: the sum of two signals generated
by linear autoregressive systems cannot be modeled efficiently with one autoregressive
model unless its order is infinite \cite{granger1976time}. This suggests that to reproduce both the speech 
and the additive signal with high quality a significantly larger model may be needed. 
Resisting the urge to increase the model size we instead performed experiments
to find out whether there is a better training and inference strategy to improve robustness 
to background noise, without changing the network configuration.

To establish best-practices for handling noise in a neural vocoder system, we run two sets of tests. 
The first uses large models with no reduction of bit rate in the inputs. 
This allows us to understand how the systems respond to noise in the best case scenario, without worry for the 
quantization schemes or model pruning techniques used. We find that placing a denoiser in front of a system 
trained on a large database of clean speech works best.

To evaluate real-world, end-to-end performance, we run an additional listening test using conversational 
speech at varying signal-to-noise ratios (SNR), with both the vocoder and denoiser models optimized for on-device performance.
In this second test we also marginalize by noise SNR and type of noise to demonstrate
that the combination of a pruned ConvTASNet and WaveRNN system works well in all circumstances.

\section{Noise Handling Strategies}

In deep learning it is common practice to apply augmentations to training data to increase the
range of conditions familiar to a model \cite{salamonaugment}. In particular, the artificial addition of noise signals is a common and 
effective step in audio event classification. This allows the model to train on a wider variety of 
realistic scenarios given a relatively clean set of ground truth data. In audio event classification,
the event labels are unchanged by the augmentations, and the goal is to train a system which is invariant under
the full set of augmentations.
Augmentations can be static (in which a new static dataset is used for training) or dynamic 
(in which augmentations are applied `on-the-fly' as new training examples are consumed). A dataset with 
dynamic augmentations is effectively infinite, though one still requires a base dataset with sufficient
variation to capture the full variety of (clean) signals to be modeled.

The neural vocoder has two sets of inputs during training: 
The aligned conditioning vectors and the teacher-forced autoregressive input signal.
The conditioning should match what is available at inference time (e.g., noisy melspectra), and the 
autoregressive input is what we measure loss against, and, therefore, what we train the model to produce. 
Observe that these do not need to be derived from the same input signal: 
In particular, if the input conditioning vectors are calculated from noise-augmented clean speech, 
we can use the raw clean speech as the teacher-forced training target. 
The result is a system which learns to produce denoised audio from noisy conditioning: 
This is similar to how augmentation is used in classification problems. The result is (hopefully) a model 
which is close to invariant under the addition of noise.

As an alternative, we can include a denoising model in the encoder.
During inference, we do not have access to the underlying clean speech, but we can apply a denoiser to
push the conditioning closer to the speech manifold.  A `perfect' denoiser would then allow a model trained
solely on clean speech to perform well, since all interfering noise has been removed. In reality, no denoiser
is perfect, and will miss some noise and introduce artifacts. Preprocessing with a denoiser is known to work
well with classical low bit rate vocoding systems \cite{wang2002melpe}, leading us to believe that a denoiser could help with 
a neural vocoder as well.

In this paper, we denote different training regimes as X2Y, where X describes the conditioning input and Y
describes the autoregressive input. The regimes we consider are:

\begin{itemize}
\item c2c: `Clean-to-Clean:' Trained with clean, studio recorded audio for both the conditioning and the autoregressive inputs.
\item n2n: `Noisy-to-Noisy:' Trained with noisy inputs, using both larger, noisier speech databases and dynamic noise 
    augmentations. This noisy data is used both for conditioning and the autoregressive inputs.
\item n2c: `Noisy-to-Clean:' Trained with noise-augmented speech from a studio-recorded database. The noise-augmented speech is
    used to compute conditioning inputs, and the original, unaugmented speech is used as the training target.
\item dc2c: `Denoised Clean-to-Clean:' The same training regime as c2c, but applies a denoiser during inference.
\item dn2n: `Denoised Noisy-to-Noisy:' The same training regime as n2n, but applies a denoiser during inference.
\end{itemize}

\section{Neural Vocoder Architecture}

In this section we describe the architecture of the neural vocoder. The parameter settings of the scheme are provided in section \ref{s:vocoderconfig}.

The vocoder consists of an encoder and decoder. The encoder simply converts the input signal to 
log melspectra (e.g., \cite{o1987speech}). The objective of the decoder is to turn these melspectra back into a high-quality speech waveform.

The overall structure of the decoder is similar to WaveRNN, with some changes which result in a leaner model, suitable for 
subsequent deployment to low-resource environments. To summarize the differences, we use a single-pass GRU, predict output 
samples using a mixture of logistics distribution, and predict $M$ frequency-banded samples at a time. (All described in full below.)

The decoder first consumes the log melspectra with a \textit{conditioning stack}, consisting of:
\begin{enumerate}
    \item An input 1D convolution (which is non-causal, allowing a fixed amount of delay based on the conditioning frame size),
    \item Three dilated causal 1D convolutions (allowing a large receptive field over the past),
    \item Three transpose convolutions, which upsample to narrow the gap between the input conditioning rate and the vocoder's output sample rate, and
    \item A final tiled upsampling so that the final output is at the vocoder's output sample rate exactly.
\end{enumerate}

The autoregressive network consists of a multi-band WaveGRU, which is based on gated recurring units (GRU) \cite{chung2014empirical}.
We split the target audio (with sample rate $S$) using a cascade of Quadrature Mirror Filters, dividing the signal evenly into $M=2^k$ frequency bands. 
Similar to \cite{okamotoSplitMoL}, this allows the system to predict $M$ samples at a time, greatly reducing the computational load
and increasing the effective receptive field resulting in a slight quality improvement.

Thus, for our $M$-band WaveGRU, $M$ samples are generated simultaneously at an update rate of $S/M$ Hz, one sample for each frequency band. 
For each update, the state of the GRU network is projected onto an $M \times K\times 3$ dimensional space that defines $M$ parameter sets, 
each set corresponding to a mixture of logistics distribution with $K$ mixture components for a particular frequency band. 
A sample for each band is then drawn by first selecting the mixture component (a logistics distribution) according to its probability and
then drawing the sample from this logistics distribution by transforming a sample from a uniform distribution \cite{salimans2017pixelcnn++}. 
For each set of $M$ output samples a synthesis filter-bank produces $M$ subsequent time-domain samples, which results in an output with sampling rate $S$ Hz.

The input to the WaveGRU consists of the sum of autoregressive and conditioning components. The autoregressive component is a projection of the last step's $M$ frequency-band samples projected onto a vector of the dimensionality of the WaveGRU state. The second component is the output of the conditioning stack (which has the same dimensionality as the WaveGRU state).

The training of the WaveGRU network and the conditioning stack is performed simultaneously using teacher forcing. That is, the past signal samples that are provided as input to the GRU are ground-truth signal samples. The training objective is maximizing the log likelihood (cross entropy) of the ground truth samples.

\begin{figure}[t]
  \centering
  \includegraphics[width=0.2\textwidth]{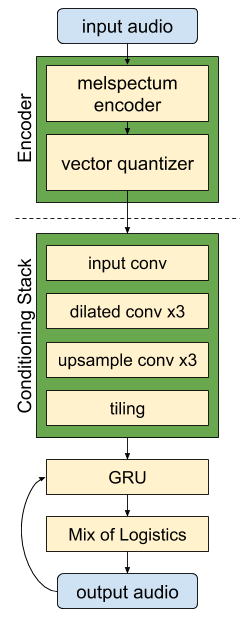}
  \caption{WaveGRU Neural Vocoder architecture.} 
  \label{fig:wavegru}
\end{figure}

\subsection{Vocoder Configuration}
\label{s:vocoderconfig}

The neural vocoder operates on 160-dimensional log melspectra computed from 80~ms windows at an update rate of 25~Hz. The system uses four frequency bands, such that the overall update rate of the WaveGRU system is 4~kHz. The conditioning stack uses 512 hidden states and a single frame (40~ms) of lookahead.  The dilated convolutional layers have kernel size two, and dilation of one, two and four respectively. Each of the three upsampling transpose convolutions doubles the rate, so that the output of the third upsampling layer is at 200~Hz. This is then tiled to match the GRU rate. The GRU state is 1024-dimensional, and eight mixture-of-logistics components are used for each output sample distribution.

For training the models we used speech from the publicly available sets WSJ0 \cite{WSJdatabase} 
and LibriTTS \cite{LibriTTS}, as well as Google proprietary TTS recordings of English speech.
These were mixed with additive noise from Freesound \cite{Font13} and a set of recordings
captured in a variety of environments, including busy streets, caf\'es and offices. 
During training of n2n and n2c models, noise samples are dynamically mixed into training samples 
with a random SNR chosen uniformly between 1~dB and 40~dB.

\subsection{On-Device Vocoder Configuration}

For the second listening test, we use a model optimized for on-device performance with a low bit rate. We apply a Karhunen-Lo\`eve transform (KLT) to each melspectrum, and then apply vector quantization to achieve a 3~kbps rate. Meanwhile, the model is pruned to 92\% sparsity using iterative magnitude pruning \cite{zhuprune} in most layers. We use 4x4 structured sparse blocks to allow fast inference using SIMD instructions \cite{kalchbrenner2018efficient}.  For the main GRU layer, we use a fixed block-diagonal sparsity pattern with 16 blocks for each of the three GRU matrices, corresponding to 93.75\% sparsity. We find this has no impact on output quality relative to magnitude pruning, and greatly improves training speed.

\section{Denoisers}

In this section we describe the architecture of the ConvTASNet denoiser\cite{luo2019conv}. We use two different denoisers. For the main MOS listening tests, we use a TDCN++ architecture, as configured in Appendix A in \cite{wisdom2020unsupervised}. This is a very high quality model, but is non-causal. Thus, it provides an upper-bound on real world quality, though the same model has been shown to work well with lower latencies \cite{sonning2020performance}.

Based on good experimental results, we then developed a causal ConvTASNet model.  Similar to the WaveGRU neural vocoder, the architecture is modified to minimize complexity when deploying to mobile devices. The parameter settings of the scheme are provided in section \ref{s:tasnetconfig}.

As with the original ConvTASNet, we use a learned filterbank $F$ with stride $H$ to transform the signal to sample rate $S/H$. The \textit{mask network} then generates sigmoid masks which are applied to the filterbanked signal. A learned transpose filterbank $G^T$ then transforms the masked signal back to the time domain.

The mask network has a separate learned filterbank with $F'$ filters and matching stride $H$. Unlike the original ConvTASNet, we remove all layer-wise normalizations, to preserve causality. We also use causal dilated convolutions and depth-wise convolutions. We allow a fixed amount of look-ahead by introducing a delay between the generated masks and the filtered mixture. As in the original ConvTASNet, we use depth-wise convolutional blocks, consisting of an `input' inverted bottleneck convolution (kernel size 1, increasing the number of channels), a depth-wise convolution, and an `output' bottleneck convolution (also with kernel size 1, decreasing the number of channels). Skip connections combine the input and output of each block.

\subsection{On-Device ConvTASNet Configuration}
\label{s:tasnetconfig}

\begin{figure}[t]
  \centering
  \includegraphics[width=0.3\textwidth]{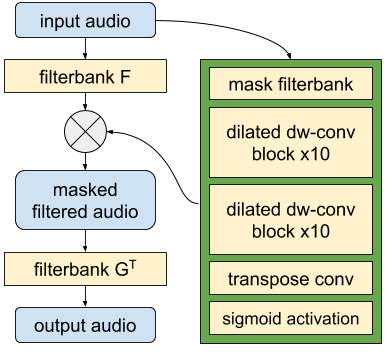}
  \caption{Convolutional TASNet architecture.} 
  \label{fig:convtasnet}
\end{figure}

The on-device model consumes audio sampled at $S=16~kHz$. The learned filterbank consists of 256 filters, with a window of 4~ms and step size of 1~ms. 

The mask network's input filterbank is the same, but with 128 filters. We use two `repeats' of ten depth-wise convolutional blocks. The depth-wise convolutions have kernel size 3 and dilation $2^{k \bmod d}$, where $k$ is the block number and $d=10$ provides a sawtooth pattern to the dilations.  The inter-block hidden size is 128, and 256 channels are used within the depth-wise convolutional blocks. Finally, a transpose convolutional layer with kernel size 3 combines outputs from adjacent time steps and matches the depth of the filterbank. A sigmoid activation is then applied to get the final masks.

The model is pruned to 95\% using iterative magnitude pruning, just as we did for the vocoder, reducing the number of parameters from 1.5M to 140k. Pruning is not applied to the input or output filterbanks or the depth-wise convolutional layers (which constitute only about 1\% of the total weights).  The unpruned model achieves a scale-invariant SNR-improvement (SI-SNRi) of 12~dB on held-out evaluation data, and the pruned model achieves 9.8~dB SI-SNRi. The pruned model can run at about 3x real-time on a single thread on a Pixel3 phone. More recent models are even smaller, and runs reliably in real time alongside the neural vocoder on the Pixel3.

\section{Experiments and discussion}
\label{s:experiments}

\begin{table*}[ht]
\centering
\caption{On-Device Models Test MOS Results. Bold entries indicate that the 95\% confidence intervals do not overlap. \label{t:table}}
\begin{tabular}{|l|ccc|ccc|ccc|ccc|}
\hline
                && All SNRs &&& 10dB SNR &&& 5dB SNR &&& 1dB SNR &\\
System          & All     & Bbl    & Amb     & All     & Bbl    & Amb     & All     & Bbl    & Amb     & All     & Bbl    & Amb     \\\hline
Reference       & 2.96    & 2.97   & 2.95    & 3.27    & 3.23   & 3.31    & 3.02    & 3.07   & 2.97    & 2.6     & 2.6    & 2.59    \\ 
TASNet          & 2.70    & 2.53   & 2.87    & 3.12    & 3.01   & 3.23    & 2.69    & 2.48   & 2.91    & 2.29    & 2.12   & 2.47    \\\hline
dn2n            & \textbf{2.18}    & \textbf{2.04}   & \textbf{2.32}    & 2.40    & 2.34   & 2.46    & \textbf{2.19}    & 1.97   & \textbf{2.41}    & \textbf{1.96 }   & \textbf{1.82}   & \textbf{2.09}    \\ 
n2n             & 1.87    & 1.83   & 1.91    & 2.21    & 2.07   & 2.36    & 1.93    & 1.93   & 1.92    & 1.46    & 1.48   & 1.44    \\ 
\hline
\end{tabular}
\label{table:acidtable}
\end{table*}

\subsection{Noise Handling Strategies Listening Test}

For our first experiment, we train unpruned vocoder models under the c2c, n2n, and n2c regimes. We also include dc2c samples, in which a large, non-causal ConvTASNet is applied to the inputs before they are fed to the c2c model. All models in this experiment use unquantized conditioning features, to study modeling and reproduction aspects of the synthesis in generative speech coding while bypassing the quantization aspects of the conditioning features.

\begin{figure}[t]
  \centering
  \includegraphics[width=0.46\textwidth]{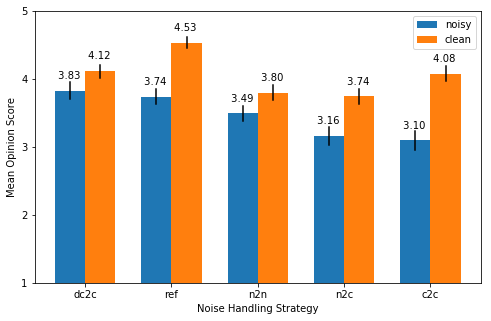}
  \caption{Mean opinion scores from the Noise Handling Strategies Listening Test. The 
    vertical bars indicate 95\% confidence intervals.} 
  \label{fig:MOS}
\end{figure}

Subjective evaluation was carried out through a crowdsourced MOS listening test, selecting $15$ clean and $15$ noisy utterances containing both male and female speakers from the VCTK dataset 
\cite{valentini2017noisy}. Each utterance had $20$ naive listeners (which could be different from utterance to utterance)
rating the quality on a scale from 1 to 5. The results are given in Fig.~\ref{fig:MOS}. 

The baseline c2c is, as expected, performing the best for clean speech but is also the worst performer in noisy speech. 
Qualitatively, the c2c model produces choppy-sounding output in regions with steady background noise. 
It also produces babbling when transient noises are present.
Using noisy features and target (n2n) improves the quality in noisy speech, but at great expense of quality in 
clean speech. 

With n2c the quality does not significantly improve for noisy speech. We find that it occasionally drops phonemes, especially `noisy' fricatives at the beginning of a word. Having access to only a single melspectrum frame of lookahead likely makes it difficult to determine whether a noisy frame is an actual speech sound or a transient background noise.

The denoised setup is the overall better system. For clean speech it has statistically indifferent 
performance from c2c, indicating that the denoiser is quite transparent in clean speech. With noisy speech it is also 
the best setup, with a mean MOS somewhat higher than the reference noisy speech. 

The conclusion from these experiments is thus that using noisy features in the training will improve 
the performance in noisy background, but the trade-off will be inferior performance in clean conditions. 
Instead, we recommend adding a speech enhancer to obtain denoised features and use clean speech as teacher-forced target
during training.   

\subsection{On-Device Models Test}

For our second experiment, we check that the denoising setup is still superior in situations closer to a real-world deployment. In this experiment, we use the pruned vocoder with melspectrum features quantized to 3~kbps, and the pruned ConvTASNet variant described above. We wish to demonstrate that inclusion of the denoiser improves a current-best n2n system, and thus use a single n2n vocoder model for both the n2n and dn2n cases.

For subjective evaluation, we used another crowdsourced MOS listening test. For this test, we use the Hispanic-English Database\cite{hispengdatabase}, which contains spontaneous, conversational speech from 22 speakers. We composed an evaluation set of twelve 10-second audio segments, eight with an isolated speaker, and two with cross-talking speakers. We added randomly selected noise samples at 1, 5 and 10 dB SNR, to produce a total of 36 evaluation segments in the test. Each item was evaluated by 30 listeners. Noise samples are either `ambient' noise (e.g., cars passing) or `babble' noise consisting of mixed background talking (e.g., background chatter in a cafe), as babble noise is a weakness of the ConvTASNet system.

Results are reported in Table \ref{table:acidtable}, both overall and marginalized by SNR and type of noise (babble vs ambient).  In summary, the dn2n system has a higher mean MOS score overall and in all marginalizations by SNR and type of noise. A 95\% confidence interval was computed for each bucket; entries where the confidence intervals did not overlap are indicated in boldface. In particular, the dn2n system is significantly better overall and at 5dB and 1dB SNR.

We also report results for the pruned ConvTASNet in isolation, and find that it does not improve on the reference in any case, and can observe that it performs a bit worse on babble noise. On listening, we find that the pruned ConvTASNet occasionally has a `scratchiness' in its output, especially at lower SNRs. Curiously, this scratchiness is largely removed in the output of the dn2n model: The artifacts may be masked by the quantized melspectrum transformation, or may be sounds that the model never saw in the training data, and are therefore smoothed away.

\section{Conclusions}

In this paper, we examined three strategies for handling noise with a neural vocoder: Adding noisy training data, training the vocoder to act as a denoiser, and adding an additional ConvTASNet denoiser in the encoder. Training on noisy data and introducing a denoiser to the encoder both worked well, though the denoiser gave the best quality. We also demonstrated that a heavily pruned ConvTASNet works well in conjunction with the neural vocoder in on-device conditions: in conversational speech with varying levels of background noise, using low biw-rate features.

\bibliographystyle{IEEEbib}
\bibliography{lyra}

\begin{thebibliography}{10}

\bibitem{oord2016wavenet}
A.~van~den Oord, S.~Dieleman, H.~Zen, K.~Simonyan, O.~Vinyals, A.~Graves,
  N.~Kalchbrenner, A.~Senior, and K.~Kavukcuoglu,
\newblock ``Wavenet: A generative model for raw audio,''
\newblock {\em arXiv preprint arXiv:1609.03499}, 2016.

\bibitem{kleijn2018wavenet}
W.~B. Kleijn, F.~S.~C. Lim, A.~Luebs, J.~Skoglund, F.~Stimberg, Q.~Wang, and
  T.~C. Walters,
\newblock ``{WaveNet} based low rate speech coding,''
\newblock in {\em 2018 IEEE International Conference on Acoustics, Speech and
  Signal Processing}, 2018, pp. 676--680.

\bibitem{garbacea2019low}
C.~G{\^a}rbacea, A.~van~den Oord, Y.~Li, F.~S.~C. Lim, A.~Luebs, O.~Vinyals,
  and T.~C. Walters,
\newblock ``Low bit-rate speech coding with {VQ-VAE} and a {WaveNet} decoder,''
\newblock in {\em 2019 IEEE International Conference on Acoustics, Speech and
  Signal Processing}, 2019, pp. 735--739.

\bibitem{klejsa2019high}
J.~Klejsa, P.~Hedelin, C.~Zhou, R.~Fejgin, and L.~Villemoes,
\newblock ``High-quality speech coding with {SampleRNN},''
\newblock in {\em 2019 IEEE International Conference on Acoustics, Speech and
  Signal Processing}, 2019, pp. 7155--7159.

\bibitem{mehri2016samplernn}
S.~Mehri, K.~Kumar, I.~Gulrajani, R.~Kumar, S.~Jain, J.~Sotelo, A.~Courville,
  and Y.~Bengio,
\newblock ``{SampleRNN}: An unconditional end-to-end neural audio generation
  model,''
\newblock {\em arXiv preprint arXiv:1612.07837}, 2016.

\bibitem{valin2019lpcnetcoded}
J.-M. Valin and J.~Skoglund,
\newblock ``A real-time wideband neural vocoder at $1.6$~kb/s using {LPCNet},''
\newblock in {\em Proc. Interspeech 2019}, 2019.

\bibitem{kalchbrenner2018efficient}
N.~Kalchbrenner, E.~Elsen, K.~Simonyan, S.~Noury, N.~Casagrande, E.~Lockhart,
  F.~Stimberg, A.~van~den Oord, S.~Dieleman, and K.~Kavukcuoglu,
\newblock ``Efficient neural audio synthesis,''
\newblock in {\em Proc. 35th International Conference on Machine Learning},
  Jennifer Dy and Andreas Krause, Eds. 2018, vol.~80 of {\em Proceedings of
  Machine Learning Research}, pp. 2410--2419, PMLR.

\bibitem{granger1976time}
C.~W.~J. Granger and M.~J. Morris,
\newblock ``Time series modelling and interpretation,''
\newblock {\em Journal of the Royal Statistical Society: Series A (General)},
  vol. 139, no. 2, pp. 246--257, 1976.

\bibitem{salamonaugment}
J.~{Salamon} and J.~P. {Bello},
\newblock ``Deep convolutional neural networks and data augmentation for
  environmental sound classification,''
\newblock {\em IEEE Signal Processing Letters}, vol. 24, no. 3, pp. 279--283,
  2017.

\bibitem{wang2002melpe}
T.~Wang, K.~Koishida, V.~Cuperman, A.~Gersho, and J.~S. Collura,
\newblock ``A 1200/2400 bps coding suite based on {MELP},''
\newblock in {\em 2002 IEEE Speech Coding Workshop}, 2002, pp. 90--92.

\bibitem{o1987speech}
D.~{O'Shaughnessy},
\newblock {\em Speech Communications: Human And Machine (IEEE)},
\newblock Universities press, 1987.

\bibitem{chung2014empirical}
J.~Chung, C.~Gulcehre, K.~Cho, and Y.~Bengio,
\newblock ``Empirical evaluation of gated recurrent neural networks on sequence
  modeling,''
\newblock {\em arXiv preprint arXiv:1412.3555}, 2014.

\bibitem{okamotoSplitMoL}
T.~{Okamoto}, K.~{Tachibana}, T.~{Toda}, Y.~{Shiga}, and H.~{Kawai},
\newblock ``An investigation of subband wavenet vocoder covering entire audible
  frequency range with limited acoustic features,''
\newblock in {\em 2018 IEEE International Conference on Acoustics, Speech and
  Signal Processing}, 2018, pp. 5654--5658.

\bibitem{salimans2017pixelcnn++}
T.~Salimans, A.~Karpathy, X.~Chen, and D.~P. Kingma,
\newblock ``{PixelCNN++}: Improving the {PixelCNN} with discretized logistic
  mixture likelihood and other modifications,''
\newblock {\em arXiv preprint arXiv:1701.05517}, 2017.

\bibitem{WSJdatabase}
{Eugene C. et al.},
\newblock ``{BLLIP} 1987-89 {WSJ} corpus release 1 ldc2000t43 web download,''
\newblock 2000.

\bibitem{LibriTTS}
H.~Zen, V.~Dang, R.~Clark, Y.~Zhang, R.~J. Weiss, Y.~Jia, Z.~Chen, and Y.~Wu,
\newblock ``{LibriTTS}: A corpus derived from {LibriSpeech} for
  text-to-speech,''
\newblock {\em arXiv}, 2019.

\bibitem{Font13}
F.~Font, G.~Roma, and X.~Serra,
\newblock ``Freesound technical demo,''
\newblock in {\em Proc. 21st ACM Int. Conf. Multimedia, Barcelona, Spain},
  2013, p. 411–412.

\bibitem{zhuprune}
M.~Zhu and S.~Gupta,
\newblock ``To prune, or not to prune: exploring the efficacy of pruning for
  model compression,''
\newblock {\em arXiv preprint arXiv:1710.01878}, 2017.

\bibitem{luo2019conv}
Y.~Luo and N.~Mesgarani,
\newblock ``{Conv-TasNet}: Surpassing ideal time--frequency magnitude masking
  for speech separation,''
\newblock {\em IEEE/ACM Transactions on Audio, Speech, and Language
  Processing}, vol. 27, no. 8, pp. 1256--1266, 2019.

\bibitem{wisdom2020unsupervised}
S.~Wisdom, E.~Tzinis, H.~Erdogan, R.~J. Weiss, K.~Wilson, and John Hershey,
\newblock ``Unsupervised sound separation using mixture invariant training,''
\newblock {\em Advances in Neural Information Processing Systems}, vol. 33,
  2020.

\bibitem{sonning2020performance}
S.~Sonning, C.~Sch{\"u}ldt, H.~Erdogan, and S.~Wisdom,
\newblock ``Performance study of a convolutional time-domain audio separation
  network for real-time speech denoising,''
\newblock in {\em 2020 IEEE International Conference on Acoustics, Speech and
  Signal Processing}, 2020, pp. 831--835.

\bibitem{valentini2017noisy}
C.~Valentini-Botinhao,
\newblock ``Noisy speech database for training speech enhancement algorithms
  and {TTS} models,''
\newblock {\em University of Edinburgh. School of Informatics. Centre for
  Speech Technology Research (CSTR)}, 2017.

\bibitem{hispengdatabase}
{W. Byrne, et al.},
\newblock ``{Hispanic-English} database {LDC2014S05} web download,''
\newblock 2014.

\end{thebibliography}
\end{document}